# Comment on "Activation of Visual Pigments by Light and Heat"[*Science* 332, 1307-312 (2011)]

V. Salari[1,2], F. Scholkmann[3], F. Shahbazi[1], I. Bokkon[4], J. Tuszynski[5]

[1]Department of Physics, Isfahan University of Technology, Isfahan 84156-83111, Iran

[2]Foundations of Physics Group, Institute for Research in Fundamental Sciences (IPM), Tehran 19395-5531, Iran

[3]University Hospital Zurich, Biomedical Optics Research Laboratory, Division of Neonatology, 8091 Zurich, Switzerland

[4]Psychoszomatic OutPatient Department of the National Center for Spinal Disorders, Hungary

[5]Department of Physics, University of Alberta, T6G 2J1, Edmonton, AB, Canada

**Abstract**:

It is known that the Arrhenius equation, based on the Boltzmann distribution, can model only a part (e.g. half of the activation energy) for retinal discrete dark noise observed for vertebrate rod and cone pigments. Luo *et al* (Science, 332, 1307-312, 2011) presented a new approach to explain this discrepancy by showing that applying the Hinshelwood distribution instead the Boltzmann distribution in the Arrhenius equation solves the problem successfully. However, a careful reanalysis of the methodology and results shows that the approach of Luo *et al* is questionable and the results found do not solve the problem completely.

**One Sentence Summary:** Retinal discrete dark noise cannot be completely explained by thermal activation based on the approach of Luo *et al*.

The application of the Arrhenius equation, which is based on the simple Boltzmann distribution, to model the temperature dependence of the dark events results in the fact that the predicted thermal activation energy is only being about half of the photo-isomerization activation energy measured experimentally (*1,2,3*). This leads to the conclusion that the molecular pathway due to spontaneous thermal activation is different from that due to photo-activation. Recently, the use of the Boltzmann distribution has been debated (*4,5*) and the idea has been put forward by Luo *et al* (*4*) that due to thermal activation of the low energy vibrational modes, the Hinshelwood distribution should be used instead of the Boltzmann distribution. Luo *et al* determined the number of vibrational modes (*m*) to be 45 in order to fill the gap between the thermal activation energy obtained from the Arrhenius analysis and the activation energy caused by light.

After carefully reviewing the approach of Luo *et al* we come to the conclusion that there are three shortcomings of this approach, which question to validity of explaining the dark noise of rods and cones by only assuming a thermal activation energy process. Our arguments were as follows:

(1) It has to be noted that the application of the Hinshelwood distribution to model *one molecule* is only valid in the classical limit where the thermal energy scale is much larger than the energy level spacing ($\varepsilon$) of the quadratic modes of the molecule (i.e. $kT \gg \varepsilon$, with $k$ is the Boltzmann constant, $T$ is the absolute temperature). Hence, assuming that the room temperature at which the thermal energy is about *25* meV, there must exist many modes with much less energies than this value. However, the opposite is true since the resonance Raman excitation of rhodopsin reveals that the Raman lines corresponds to several tens of modes with energies varying from 98 cm$^{-1}$ to 1655 cm$^{-1}$ (corresponding to ~10 to ~200 meV, respectively) which are in order or larger than the scale of the thermal energy (*6, 7, 8*). Moreover, Luo *et al* obtained *45* modes were found to have equal energy values, $kT$ ("[…] each vibrational mode of the molecule contributing a nominal energy of $kT$"(*4*)) in which the *45* modes all are activated and each energy mode has exactly the same energy as the thermal energy. As a conclusion, the *equipartition theorem* (*9*) cannot be applied for these modes; hence the application of the Hinshelwood distribution to model the dark noise of photoreceptors is questionable.

(2) Even if we agree that the Hinshelwood distribution is applicable for photoreceptors then the methodology and the obtained results by Luo *et al* can be questioned. The authors determined that the number of modes (i.e. *m* = 45) is generally valid for the all values of $\lambda_{max}$ (see **Fig 4C'**[1] and **Fig S8'**) while this value is obtained only via a simple equation (*4, 10*) for *Bufo* red rhodopsin with $\lambda_{max}$ = 500 nm based on the apparent thermal activation energy of 21.9 kcal/mol. If we consider mouse rhodopsin with the same $\lambda_{max}$ = 500 nm and use the apparent thermal activation energy of 14.54 kcal/mol (obtained from **Fig. S4C'**) we find *m* = 58 (*10*) according to the methodology used by Luo *et al*. This indicates that the statement of Luo *et al* regarding the general validity of the parameter value m = 45 is not supported by experimental findings. To show discrepancy more clearly, the rate constant diagrams based on *m* = 45 and our obtained *m* values (i.e. *m=49* for rod cells and *m=42* for cone cells*)* based on the fitting method *(10)* are compared with the experimental data for different *rod* and *cone* cells *(10, 3)* in **Fig. 1**. The results indicate that *m=45* is not an exclusive value and has a significant deviation relative to the experimental data. Moreover, our obtained pre-exponential factors (*A*) deviate from the *A* value used by Luo *et al* (see **Table S4'**) which was obtained by the authors by simple averaging and not by fitting, which is imprecise as well.
If we apply the average of *m=42* and *m=49* as *m=45* with a single *A* value for a combined datasets I and II for both rod and cone cells then the amount of deviation from experimental data will be very large. To check this high deviation see below the comment 3 about predictions of the Table 1'. As a result, rod and cone cells should be investigated separately with different *m* values.

---
[1] The primed numbers for tables and figures refer to the paper of Luo et al (*4*).

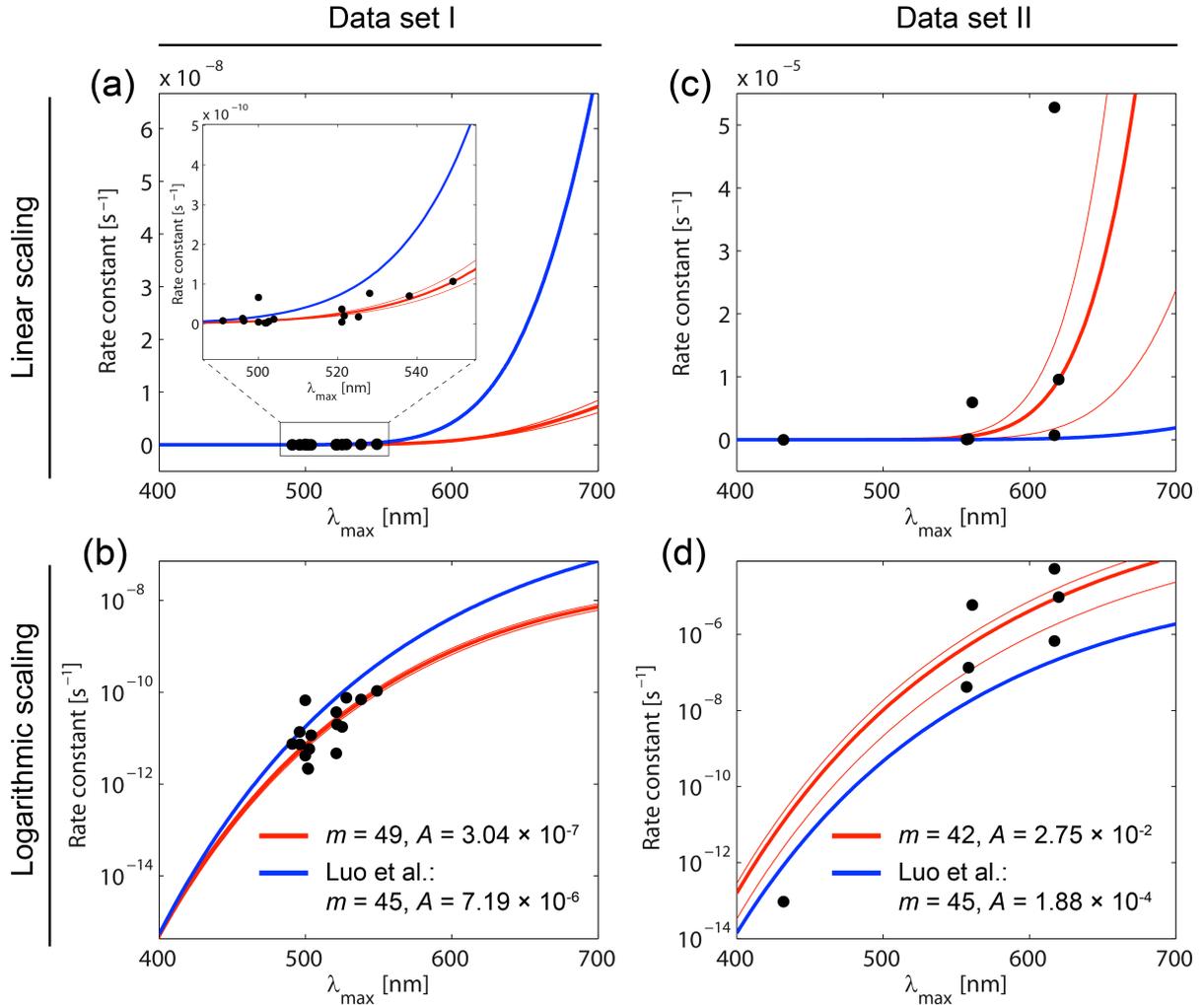

**Fig. 1**. Fitted functions (rate constant vs. $\lambda_{max}$) according to Equation 1' of (*4*) with optimal *m*-values (red) and *m* = 45 (blue) as predicted by Luo *et al* (*4*). Shown are the results with a linear (a, b) and logarithmic scaling (c, d). (a, b): data set I (rhodopsins), (c, d): data set II (cone pigments) *(10)*.

(3) Another problem in the paper of Luo *et al* appears in their predictions given in **Table 1'**. There, the authors claimed that the ratios of rate constants, *k*, are the ratios of their distribution functions, $f_{\geq E_a^T}$ ("[…] We began with *A* being the same for the all pigments […] thus the predicted thermal rate ratio between two pigments is simply their $f_{\geq E_a^T}$ ratio"(*4*)). However, the pre-exponential factor *A* varies with *m* (e.g. see the caption of **FigS8'**) and it varies for different pigments, so it is unfortunately erroneous to compare the distribution ratios (as predicted rate constant ratios) with the measured rate-constant ratios in **Table1'** while the *A* values are not equal even for the same number of modes (i.e. *m* = 45) for cone and rod cells (see **Table S4'** and also the average *A* and *SD* values for rod cells). Moreover, these pre-exponential factors of **Table S4'** are obtained directly from the measured rate constants themselves which causes an unfair comparison. The authors have mentioned that there is about 26-fold difference between *A* values of rods and cones. To check this claim, we compared these ratios for different samples. The results are shown in **Table1** in which large discrepancies between theory and experiment can be recognized, indicating that the distribution ratios are not equal to the rate constant ratios.

**Table 1.** Comparison between theoretical predictions offered by Luo *et al* (for *m* = 45) and the measurements of rate constants of visual pigments (*10*). Even for similar $\lambda_{max}$ values of rods (*Bufo* and mouse) and cones (human and turtle), 16 and 83 fold difference appeared respectively. For other comparisons between rods and cones the differences are very large numbers which indicates that predicted and measured rate constants are not comparable.

| Pigment | $\lambda_{max}$ (nm) | $E_a$(kcal mol$^{-1}$) | $f_{\geq E_a^T}$ | Predicted rate constant ratio (Luo et al.'s approach) | Measured rate constant (s$^{-1}$) | Measured rate constant ratio |
|---|---|---|---|---|---|---|
| *Bufo* rhodopsin | 500 | 48.03 | 3.65×10$^{-6}$ | 1 | 4.18×10$^{-12}$ | $\frac{1}{16}$ |
| mouse rhodopsin | 500 | 48.03 | 3.65×10$^{-6}$ | | 6.64×10$^{-11}$ | |
| human red cone | 617 | 38.93 | 2.44×10$^{-3}$ | 1 | 6.70×10$^{-7}$ | $\frac{1}{83}$ |
| Turtle (Trachemysscriptaelegans) L-cone | 617 | 38.93 | 2.44×10$^{-3}$ | | 5.28×10$^{-5}$ | |
| Larval tiger salamander (Ambystomatigrinum) rod | 521 | 46.10 | 1.67×10$^{-5}$ | $\frac{1}{147}$ | 4.69×10$^{-12}$ | $\frac{1}{11261261}$ |
| Turtle (Trachemysscriptaelegans) L-cone | 617 | 38.93 | 2.44×10$^{-3}$ | | 5.28×10$^{-5}$ | |
| Sturgeon (Acipenserbaeri) rods | 549 | 43.75 | 7.45×10$^{-5}$ | $\frac{1}{2}$ | 1.07×10$^{-10}$ | $\frac{1}{55555}$ |
| Macaque (Macacafascicularis) L-cone | 561 | 42.82 | 1.36×10$^{-4}$ | | 5.94×10$^{-6}$ | |
| Cane toad (Bufomarinus) red rod | 503.9 | 47.67 | 4.72×10$^{-6}$ | $\frac{1}{29.4}$ | 1.17×10$^{-11}$ | $\frac{1}{526315}$ |
| Macaque (Macacafascicularis) L-cone | 561 | 42.82 | 1.36×10$^{-4}$ | | 5.94×10$^{-6}$ | |

In conclusion, careful reanalysis of the methodology and results shows that the approach of Luo *et al* is questionable. We believe that their approach for the origin of retinal discrete dark noise suffers from major deficiencies and the results obtained do not offer acceptable solutions to the problem.

**References and Notes:**

[1]   D. A. Baylor, G. Matthews, K. W. Yau, Two components of electrical dark noise in toad retinal rod outer segments. *J. Physiol.* **309**, 591-621 (1980).
[2]   A. Cooper, Energy uptake in the first step of visual excitation. *Nature* **282**, 531-533 (1979).
[3]   P. Ala-Laurila, K. Donner, A. Koskelainen, Thermal activation and photoactivation of visual pigments. *Biophys. J.* **86**, 3653-3662 (2004).
[4]   D. G. Luo, W. W. Yue, P. Ala-Laurila, K. W. Yau, Activation of visual pigments by light and heat. *Science* **332**, 1307-312 (2011).
[5]   S. Gozem, I. Schapiro, N. Ferré, M. Olivucci, The molecular mechanism of thermal noise in rod photoreceptors. *Science* **337,** 1225-1228 ( 2012).
[6]   G.R. Loppnow, R. A. Mathies, Excited-state structure and isomerization dynamics of the retinal chromophore in rhodopsin from resonance Raman intensities. *Biophys. J.* **54**, 35–43 (1988).
[7]   S.W. Lin, M. Groesbeek, I. van der Hoef, P. Verdegem, J. Lugtenburg, and R. A. Mathies, Vibrational assignment of torsional normal modes of rhodopsin: probing excited-state isomerization dynamics along the reactive C11 1⁄4 C12 torsion coordinate. *J. Phys. Chem. B.* **102**, 2787– 2806 (1998).
[8]   J.E. Kim, M. J. Tauber, and R. A. Mathies, Analysis of the mode- specific excited-state energy distribution and wavelength-dependent photoreaction quantum yield in rhodopsin. *Biophys. J.* **84**:2492–2501 (2003).
[9]   L. E. Reichl, "A Modern Course in Statistical Physics", 2nd Edition, (John Wiley & Sons (1998)).
[10]  Supplementary Materials for Comment on "Activation of Visual Pigments by Light and Heat"
[11]  R. B. Barlow, R. R. Birge, E. Kaplan, J. R. Tallent, On the molecular origin of photoreceptor noise. *Nature* **366**, 64-66 (1993).

# Supplementary Materials

## 1. Determination of the number of modes based on Luo et al's method

Luo et al. (*1*) determined the number of molecular vibrational modes (*m*) only for *Bufo* red rhodopsin and applied it (i.e., *m* = 45) to all types of photoreceptors. The *m* value for *Bufo* red rhodopsin with $\lambda_{max} = 500 nm$ is obtained based on the equation $E_a^T - E_a^{T(app)} = (m-1)RT$, where $E_a^T$ is the thermal isomerization activation energy of 48.03 kcal/mol, $E_a^{T(app)}$ the apparent thermal activation energy of 21.9 kcal/mol (*1*), *R* the universal gas constant, and *T* the absolute temperature. If we consider mouse rhodopsin with the same $\lambda_{max}$ value as the Bufo rhodopsin, i.e. $\lambda_{max} = 500 nm$, and use the apparent thermal activation energy of 14.54 kcal/mol (obtained from Fig. S4C') we find *m* = 58, which is different than the generally valid relation *m* = 45 proposed by Luo et al. In addition, a value of *m* different from 45 and 58 is obtained for $A_1$ human red cones (see **Table 1S**). In this case, the apparent thermal energy values were taken from the paper of Luo et al. (*1*). Based on the findings of our analysis we conclude that the method used by Luo et al. is not generally valid and should be improved.

**Table1S.** The pigments given in (*1*) have been revised based on the apparent thermal energy, $E_a^{T(app)}$, and the *m* value. It is seen that the $A_1$ *Bufo* rhodopsin and the $A_1$ mouse rhodopsin have similar $\lambda_{max}$ values ($\lambda_{max}$ = 500 nm) while their $E_a^{T(app)}$ were different. This causes a significant difference in the *m* values. Another *m* value is obtained for the $A_1$ human red cone, which is again different than the 'exclusive' value *m* = 45.

| Pigment | $\lambda_{max}$ [nm] | $E_a$ [kcal mol-1] | Measured rate constant [s$^{-1}$] | $E_a^{T(app)}$ [kcal mol-1] | *m* |
|---|---|---|---|---|---|
| *A1 Bufo rhodopsin* | 500 | 48.03 | 4.18 × 10-12 | 21.9 | 45 |
| A1 mouse rhodopsin | 500 | 48.03 | 6.64 × 10-11 | 14.54 | 58 |
| A2 *Xenopus* rhodopsin | 521 | 46.10 | 3.70 × 10-11 | *Not specified* | – |
| A1 human red cone | 557 | 43.12 | 4.14 × 10-8 | 14.64 | 50 |
| A2 human red cone | 617 | 38.93 | 6.70 × 10-7 | *Not specified* | – |
| A1 *Bufo* blue cone | 432 | 55.59 | 9.39 × 10-14 | *Not specified* | – |

## 2. Determination of the optimal number of modes

The optimal number of molecular vibrational modes (*m*) contributing thermal energy to the pigment's activation was found by (i) computing Equation 1 of (*1*) for *m* = 35, 36, ..., 65 (data set I, see **Table 1S**) and *m* = 42, 43, ..., 49 (data set II, see **Table 2S**), respectively; (ii) fitting the functions with the free parameter *A* to the data sets; and (iii) determining the goodness of fit by calculating the root-mean-squared error (RMSE). The function with the lowest RMSE value was then chosen as the function describing the relationship between $\lambda_{max}$ and the rate constant in the best way. A robust nonlinear least squares fitting with the least absolute residuals (LAR) method (*2*) and the Levenberg–Marquardt algorithm (LMA) (*3-4*) was used. The advantage of LAR over ordinary least squares (OLS) is that the method is more robust against deviations from the normality assumption of the data. LMA combines the advantages of gradient-descent and Gauss-Newton methods in order to determine a global minimum of a function.

The best fit is obtained for *m* = 49 (data set I, RMSE = 1.055 × 10$^{-11}$, *A* = 3.04 × 10$^{-7}$) and *m* = 42 (data set II, RMSE = 5.13× 10$^{-6}$, *A* = 2.75 × 10$^{-2}$), respectively (see **Fig. 1S**). The results obtained by Luo et al. (*1*), the functions for *m* = 45 and *A* = 7.19 × 10$^{-6}$ (rhodopsins) and *A* = 1.88 × 10$^{-4}$ (cone pigments), were plotted, too. **Fig. 1S** shows the fitted functions for the optimal *m*-values for data sets I and II. In addition to the optimal functions, the function for *m* = 45 was plotted as well as the 95% confidence bound of the fitting procedure. The functions were plotted in **Fig. 1** with a linear scale (a, b) as well as with a logarithmic one (c, d).

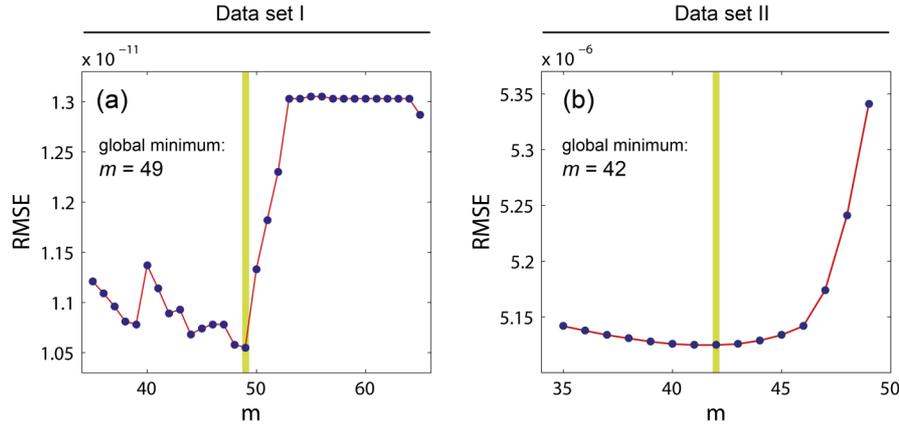

**Fig. 1S:** RMSE values for the curve fitting with Equation 1 of (*1*) performed on the data set I (a) and II (b). The global minima were highlighted as green vertical lines.

We also determined the best *m* and *A* values for the combination of data sets I, II. Therefore, the fitting was performed with LAR and LMA whereas the robust fitting option was not used since the data points showed a large amount of heteroscedasticity. The optimal values determined were *m* = 49 and *A* = 4.62 × 10$^{-3}$ (see **Fig.2S**). From this figure it is clearly seen that the separate treatment of cones and rods (i.e. data sets I and II) improves the fitting of the function.

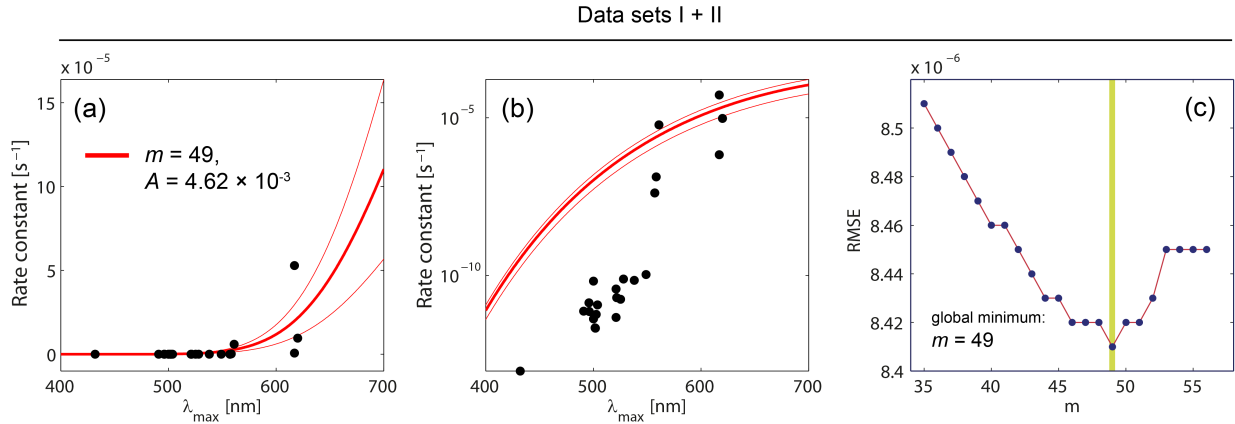

**Figure 2S:** Results of fitting Equation 1 of (*1*) to the combined data set (I and II). Optimal function with m = 49 and combined data set with linear (a) and logarithmic scaling (b). RMSE values for *m* = 35, 36, ..., 56.

**Table 1S**: Data set I consisting of rods and rhodopsins. Data are taken from Luo et al. (*1*), which are measured/obtained at 23⁰C, and from Ala-Laurila et al. (*5*) at 21⁰C. It can be simply shown that the difference between measurements at these two temperatures is trivial and does not affect the values.

| Species, type | $\lambda_{max}$ [nm] | Measured rate constant [s$^{-1}$] |
|---|---|---|
| *Bufo,* rhodopsin | 500 | 4.18×10$^{-12}$ |
| Mouse, rhodopsin | 500 | 6.64×10$^{-11}$ |
| *Xenopus,* rhodopsin | 521 | 3.70×10$^{-11}$ |
| Salamander, rhodopsin | 502 | 2.13×10$^{-12}$ |
| Salamander, rhodopsin | 528 | 7.66×10$^{-11}$ |
| Macaque (*Macacafascicularis*), rods | 491 | 7.45×10$^{-12}$ |
| Dogfish (*Scyliorhinuscanicula*), rods | 496 | 1.36×10$^{-11}$ |
| Human, rods | 496.3 | 7.30×10$^{-12}$ |
| Bullfrog (*Ranacatesbeiana*), rhodopsin | 501.7 | 2.21×10$^{-12}$ |
| Common toad (*Bufobufo*), red rods | 502.6 | 5.86×10$^{-12}$ |
| Cane toad (*Bufomarinus*), red rods | 503.9 | 1.17×10$^{-11}$ |
| Larval tiger salamander (*Ambystomatigrinum*)(A2), rods | 521 | 4.69×10$^{-12}$ |
| Clawed frog (*Xenopuslaevis*), rods | 521.6 | 2.00×10$^{-11}$ |
| Bullfrog (*Ranacatesbeiana*) porphyropsin rods | 525.2 | 1.76×10$^{-11}$ |
| Hybrid sturgeon (*Husohuso X Acipensernudiventris*) rods | 538 | 7.00×10$^{-11}$ |
| Sturgeon (*Acipenserbaeri*) rods | 549 | 1.07×10$^{-10}$ |

**Table 2S**: Data set II consisting of cone pigments. Data are taken from Luo et al. (*1*) and Ala-Laurila et al. (*5*).

| Species, type | $\lambda_{max}$ [nm] | Measured rate constant [s$^{-1}$] |
|---|---|---|
| Human, red cone | 617 | $6.70 \times 10^{-7}$ |
| Turtle (Trachemysscriptaelegans), L-cone | 617 | $5.28 \times 10^{-5}$ |
| Human, red cone | 557 | $4.14 \times 10^{-8}$ |
| *Bufo,* blue cone | 432 | $9.39 \times 10^{-14}$ |
| Salamander, cone | 557 | $4.14 \times 10^{-8}$ |
| Human, L-cone | 558.4 | $1.34 \times 10^{-7}$ |
| Macaque (*Macacafascicularis*), L-cone | 561 | $5.94 \times 10^{-6}$ |
| Larval tiger salamander (*Ambystomatigrinum*), L-cone | 620 | $9.58 \times 10^{-6}$ |

**References**
[1] D.-G. Luo, Yue, W.W. S. Yue, P. Ala-Laurila, K.-W. Yau, Activation of visual pigments by light and heat. *Science* **332**, 1307–1312 (2011).
[2] R. W. Hill and P. W. Holland, Two robust alternatives to least-squares regression. *J. Am. Stat. Assoc.* **72**(360a), 828–833 (1977).
[3] K. Levenberg, A method for the solution of certain non-linear problems in least squares. *Quart. J. Appl. Math.* **2**(2), 164–168 (1944).
[4] D. Marquardt, An algorithm for least-squares estimation of nonlinear parameters, *SIAM J. Appl. Math.* **11**(2), 431–441 (1963).
[5] P. Ala-Laurila, K. Donner, A. Koskelainen, Thermal activation and photoactivation of visual pigments. *Biophys. J.* **86**, 3653–3662 (2004).